\documentclass[reprint,aps,prd,twocolumn,notitlepage,nofootinbib,showpacs,preprintnumbers,superscriptaddress]{revtex4-1}
\usepackage[T1]{fontenc}
\usepackage[utf8]{inputenc}
\usepackage{bm}
\usepackage{amsmath}
\usepackage{amssymb}
\usepackage{esint}
\usepackage{color}
\usepackage{graphicx}
\PassOptionsToPackage{normalem}{ulem}
\usepackage{ulem}

\usepackage{times}
\usepackage{hyperref}

\begin{document}

\title{On covariant expansion of the gravitational St\"{u}ckelberg trick}

\author{Xian Gao}%
    \email[Email: ]{gao@th.phys.titech.ac.jp}
    \affiliation{%
        Department of Physics, Tokyo Institute of Technology,\\ 
        2-12-1 Ookayama, Meguro, Tokyo 152-8551, Japan}

\date{February 27, 2015}

\begin{abstract}
	A new approach to expanding the ``St\"{u}ckelbergized'' fiducial metric in a covariant manner is developed.
	The idea is to consider the curved 4-dimensional space as a codimension-one hypersurface embedded in a 5-dimensional Minkowski bulk, in which the 5-dimensional Goldstone modes can be defined as usual.
	After solving one constraint among the five 5-dimensional Goldstone modes and projecting onto the 4-dimensional hypersurface, we are able to express the ``St\"{u}ckelbergized'' fiducial metric in terms of the 4-dimensional Goldstone modes as well as 4-dimensional curvature quantities.
	We also compared the results with expressions got using the Riemann Normal Coordinates (RNC)  in Gao et al [Phys. Rev. D90, 124073 (2014)] and find that, after a simple field redefinition, results got in two approaches exactly coincide.
\end{abstract}

\maketitle

\section{Itroduction}

Attempts to explain the primordial and late time accelerating expansion of our Universe stimulate the study of theories beyond general relativity (GR) with a cosmological constant (see \cite{Clifton:2011jh,Joyce:2014kja} for reviews and \cite{Khoury:2013tda} for a short introduction).
Such theories typically contain additional degrees of freedom to the two tensor modes of GR. 
Instead of introducing extra fields by hand, one approach to these new degrees of freedom is to construct effective theories with less gauge redundancies comparing with GR. 
This can be achieved most straightforwardly by introducing a fiducial metric $\bar{f}_{\mu\nu}$, which does not change under the coordinates transformation and thus explicitly breaks the general covariance.

If $\bar{f}_{\mu\nu}$ is degenerate and has only one nonvanishing timelike eigenvector, one gets the effective field theory (EFT) of inflation \cite{Creminelli:2006xe,Cheung:2007st} and recent so-called ``theories beyond Horndeski'' \cite{Gleyzes:2014dya,Gao:2014soa}, where the time diffeomorphism is broken and thus generally an additional scalar mode arises \cite{Gleyzes:2014qga,Lin:2014jga,Gao:2014fra}.
Spatial symmetries may be broken by considering $\bar{f}_{\mu\nu}$ with nonvanishing spacelike eigenvectors.
If the number of nonvanishing spacelike eigenvectors is equal to the spatial dimension, one arrives at some sort of massive gravity theories (typically Lorentz-violating, e.g. \cite{Comelli:2013txa,Lin:2015cqa}).
A nondegenerate $\bar{f}_{\mu\nu}$ breaks all spacetime symmetries, through which a Lorentz-invariant massive gravity can be constructed \cite{deRham:2010kj} (see \cite{Hinterbichler:2011tt,deRham:2014zqa} for reviews).
In this note, we concentrate on the case of a nondegenerate $\bar{f}_{\mu\nu}$.

The fiducial metric $\bar{f}_{\mu\nu}$, whose existence breaks the general covariance explicitly, can always be thought of as the ``gauge-fixed'' version of some covariant tensor field. 
This is just the idea of gravitational St\"{u}ckelberg trick, which dates from  \cite{Green:1991pa,Siegel:1993sk} in the study of open string theory.
We may promote the fiducial metric $\bar{f}_{\mu\nu}$ as \cite{Siegel:1993sk,ArkaniHamed:2002sp}
\begin{equation}
\bar{f}_{\mu\nu} \rightarrow f_{\mu\nu}(x) \equiv \bar{f}_{ab}(\phi(x)) \frac{\partial\phi^a(x)}{\partial x^{\mu}} \frac{\partial\phi^b(x)}{\partial x^{\nu}}, \label{stueckelberg}
\end{equation}
where the ``St\"{u}ckelbergized'' fiducial metric $f_{\mu\nu}$ transforms as a tensor as long as each of the four (we are working with 4-dimensional spacetime) St\"{u}ckelberg fields $\{\phi^a\}$ transform as scalars under a general coordinates transformation of spacetime.
The fixed $\bar{f}_{\mu\nu}$ is recovered by choosing the so-called ``unitary gauge'' with $\phi^a \rightarrow \delta^a_{\mu} x^{\mu}$.
In practice, we may expand the St\"{u}ckelberg fields around the unitary gauge and concentrate on the behavior of their fluctuations \cite{ArkaniHamed:2002sp} 
	\begin{equation}
	\phi^a - \delta^a_{\mu} x^{\mu} \equiv -\delta^a_{\mu} \hat{\pi}^{\mu}. \label{naive}
	\end{equation}
When the fiducial metric $\bar{f}_{\mu\nu}$ is flat, it has been well-known that in the so-called decoupling limit (some limit of energy scales where the interactions among different types of degrees of freedom get simplified), $\hat{\pi}_{\mu}$ defined in (\ref{naive}) behaves as a spacetime vector.
In this case, we can fix a gauge in which the helicity-1 and helicity-0 parts of the graviton are encoded in $\hat{\pi}_{\mu}$ \cite{deRham:2011qq}.
It is just in this way that the Boulware-Deser ghost \cite{Boulware:1973my} can be seen most transparently  \cite{Deffayet:2005ys,Creminelli:2005qk}.
This argument, however, cannot be simply applied to a general fiducial metric $\bar{f}_{\mu\nu}$. 
First, naively plugging (\ref{naive}) into (\ref{stueckelberg}) would inevitably yield noncovariant expressions \cite{deRham:2012kf,Gao:2014ula}.
More seriously, as was well explained in  \cite{Hassan:2011vm,deRham:2011rn,deRham:2012kf}, $\hat{\pi}_\mu$ defined in (\ref{naive}) is not a vector and does not capture the helicity-1 and helicity-0 modes of the graviton correctly, either when going beyond the decoupling limit or when the fiducial metric has curvature.

This problem was systematically solved in \cite{Gao:2014ula} by employing the the Riemann Normal Coordinates (RNC), where a covariant formulation of the St\"{u}ckelberg expansion with a general fiducial metric was developed. 
A decoupling limit analysis similar to the case of a flat fiducial metric was consistently performed in \cite{Gao:2014ula}, where the helicity modes can be characterized correctly\footnote{See also \cite{Mirbabayi:2011aa,Alberte:2011ah,Fasiello:2013woa,Kugo:2014hja} for related progresses on the St\"{u}ckelberg analysis and decoupling limit of massive gravity around a general background.}.
On the other hand, when dealing with the de Sitter fiducial metric, an alternative approach to the covariant St\"{u}ckelberg expansion was developed in \cite{deRham:2012kf}.
The idea is to embed the $d$-dimensional de Sitter space into a $(d+1)$-dimensional Minkowski background, in which the Goldstone modes can be identified as in (\ref{naive}).
Then by projecting to the $d$-dimensional de Sitter space, (\ref{stueckelberg}) can be expanded in terms of the correct helicity modes in $d$ dimensions, in a covariant manner. 
The purpose of this note, is to develop this technique further, and more systematically, by considering a general fiducial metric.

This note is organized as follows.
In Sec.\ref{sec:RNC}, we briefly review the main results in \cite{Gao:2014ula} on the covariant St\"{u}ckelberg expansion based on RNC.
In Sec.\ref{sec:embed} we establish the basic formalism of embedding the $4$-dimensional curved space into a $5$-dimensional Minkowski one, and determine the ``covariant'' Goldstone modes in the 4 dimensions.
In Sec.\ref{sec:expan}, we derive the covariant expansions for the fiducial metric in terms of 4-dimensional quantities, and compare them with the corresponding results in \cite{Gao:2014ula}.
Finally we briefly summarize in Sec.\ref{sec:con}.

\section{Covariant expansion based on the Riemann normal coordinates} \label{sec:RNC}

In \cite{Gao:2014ula}, the Riemann Normal Coordinates (RNC) was employed to derive covariant expressions for the St\"{u}ckelberg expansion in the presence of a general fiducial metric\footnote{This is in fact a standard approach in order to define perturbations ``covariantly'', as in the well-known background field method \cite{AlvarezGaume:1981hn}. 
For massive gravity, RNC was also suggested in \cite{Hassan:2011vm} (footnote 5).}.
The idea is to consider a one-parameter family of diffeomorphisms of the spacetime with parameter $\lambda$:
	\begin{equation}
		\phi_{\lambda}: p \mapsto \phi_{\lambda}(p),
	\end{equation}
where $p$ is a given spacetime point.
The St\"{u}ckelberg fields at point $p$ are defined as the coordinate values of its image $\phi_{\lambda}(p)$ with $\lambda=-1$:
	\begin{equation}
		\left. \phi^{\mu}\right|_{p} \equiv \left. x^{\mu} \right|_{\phi_{-1}(p)}.
	\end{equation}
The ``covariant'' Goldstone modes $\pi^{\mu}$ are thus defined as the standard RNC's of the image point $\phi_{-1}(p)$, i.e., the tangent vector of the geodesic at point $p$ connecting $p$ and its image $\phi_{-1}(p)$:
	\begin{eqnarray}
	\left.\phi^{\mu}\right|_{p} & = & x_{0}^{\mu}-\pi^{\mu}-\frac{1}{2}\bar{\Gamma}_{\nu\rho}^{\mu}\pi^{\nu}\pi^{\rho}\nonumber \\
	&  & +\frac{1}{6}\left(\partial_{\nu}\bar{\Gamma}_{\rho\sigma}^{\mu}-2\bar{\Gamma}_{\nu\lambda}^{\mu}\bar{\Gamma}_{\rho\sigma}^{\lambda}\right)\pi^{\nu}\pi^{\rho}\pi^{\sigma}+\cdots, \label{rnc}
	\end{eqnarray}
where $x^{\mu}_0 \equiv \left.x^{\mu}\right|_{p}$, and $\bar{\Gamma}^{\mu}_{\nu\rho}$ is the Christoffel symbol associated with $\bar{f}_{\mu\nu}$.
Comparing with (\ref{naive}), (\ref{rnc}) also implies a nonlinear relation between $\hat{\pi}_{\mu}$ and $\pi_{\mu}$.

By plugging (\ref{rnc}) into (\ref{stueckelberg}) and carefully dealing with the Christoffel symbols and their derivatives, it is possible to expand the ``St\"{u}ckelbergized'' fiducial metric $f_{\mu\nu}$ covariantly in terms of $\pi_{\mu}$, $\bar{R}_{\mu\nu\rho\sigma}$ as well as  their covariant derivatives (with respect to $\bar{f}_{\mu\nu}$).
An equivalent and more convenient approach, is to evaluate
(see \cite{Gao:2014ula} for details)
	\begin{eqnarray}
	f_{\mu\nu} = e^{-\pounds_{\bm{\pi}}}\bar{f}_{\mu\nu} 
	& \equiv & \sum_{n=0}\frac{\left(-1\right)^{n}}{n!}\pounds_{\bm{\pi}}^{n}\bar{f}_{\mu\nu}\nonumber \\
	& = & \sum_{n=0}f_{\mu\nu}^{(n)\mathrm{RNC}},
	\end{eqnarray}
where $\pounds_{\bm{\pi}}^{n}\bar{f}_{\mu\nu}$ is the $n$-th order Lie derivative of $\bar{f}_{\mu\nu}$ with respect to $\pi^{\mu}$.
Straightforward calculations yield\footnote{Note in \cite{Gao:2014ula}, $f^{(n)\mathrm{RNC}}_{\mu\nu}$ were evaluated only up to the cubic order in $\pi^{\mu}$. Here we also evaluate $f^{(4)\mathrm{RNC}}_{\mu\nu}$ for late convenience.}
\begin{eqnarray}
f_{\mu\nu}^{(1)\mathrm{RNC}} & = & -2\bar{\nabla}_{(\mu}\pi_{\nu)},\label{f1_RNC}\\
f_{\mu\nu}^{(2)\mathrm{RNC}} & = & \bar{\nabla}_{\mu}\pi_{\rho}\bar{\nabla}_{\nu}\pi^{\rho}-\bar{R}_{\mu\rho\nu\sigma}\pi^{\rho}\pi^{\sigma},\label{f2_RNC}\\
f_{\mu\nu}^{(3)\mathrm{RNC}} & = & \frac{1}{3}\pi^{\alpha}\pi^{\beta}\left(\pi^{\rho}\bar{\nabla}_{\rho}\bar{R}_{\mu\alpha\nu\beta}+4\bar{\nabla}_{(\mu}\pi^{\rho}\bar{R}_{\nu)\alpha\rho\beta}\right),\label{f3_RNC}\\
f_{\mu\nu}^{(4)\mathrm{RNC}} & = & -\frac{1}{12}\left(\bar{\nabla}_{\alpha}\bar{\nabla}_{\beta}\bar{R}_{\mu\rho\nu\sigma}-4\bar{R}_{\mu\alpha\beta\lambda}\bar{R}_{\nu\rho\sigma}^{\phantom{\nu\rho\sigma}\lambda}\right)\pi^{\alpha}\pi^{\beta}\pi^{\rho}\pi^{\sigma}\nonumber \\
&  & +\frac{1}{2}\bar{\nabla}_{\rho}\bar{R}_{\alpha\beta\sigma(\mu}\bar{\nabla}_{\nu)}\pi^{\alpha}\pi^{\beta}\pi^{\rho}\pi^{\sigma}\nonumber \\
&  & -\frac{1}{3}\bar{R}_{\rho\alpha\sigma\beta}\pi^{\alpha}\pi^{\beta}\bar{\nabla}_{\mu}\pi^{\rho}\bar{\nabla}_{\nu}\pi^{\sigma}.\label{f4_RNC}
\end{eqnarray}
Please note in deriving (\ref{f2_RNC})-(\ref{f4_RNC}), $\pi^{\nu}\bar{\nabla}_{\nu}\pi^{\mu} = 0$ is used, since $\pi^{\mu}$ is the tangent vector of geodesics.

\section{St\"{u}ckelberg by embedding} \label{sec:embed}

A different approach to the covariant St\"{u}ckelberg expansion was introduced in \cite{deRham:2012kf} in the study of massive gravity on de Sitter background.
This approach is based on the observation that the $d$-dimensional de Sitter space can be embeded into a $(d+1)$-dimensional Minkowski one, in which the Goldstone modes can be identified easily as in (\ref{naive}) and the St\"{u}ckelberg expansion of the fiducial metric can be performed as usual.
Then the $d$-dimensional quantities can be got, in a covariant manner, by simply projecting the $(d+1)$-dimensional ones onto the $d$-dimensional de Sitter space.

In general, to embed an arbitrary $d$-dimensional space into a $(d+1)$-dimensional Minkowski one is not always possible.
While as the first attempt, in this note, we restrict ourselves to the subclass of 4 dimensional metrics which can be embedded (at least locally) into a 5 dimensional flat bulk.
In this case, the corresponding St\"{u}ckelbergized fiducial metric is given by
	\begin{equation}
		f_{\mu\nu} = \eta_{MN} \frac{\partial X^M}{\partial x^{\mu}} \frac{\partial X^N}{\partial x^{\nu}}, \label{f_ind}
	\end{equation}
where $\{X^M\}$ with $M=0,1,2,3,4$ are Cartesian coordinates of the 5-dimensional Minkowski space.
As we shall see, in the 5-dimensional flat bulk, the unitary gauge and the decomposition of the St\"{u}ckelberg fields can be performed in a standard manner. All the subtleties are thus in the projection from 5 dimensions to 4 dimensions.

It is convenient to introduce another set of coordinates $\{x^a\}$ for the 5-dimensional Minkowski bulk such that its metric takes the form
	\begin{equation}
		\mathrm{d}s^{2}=\mathfrak{h}_{ab}dx^{a}dx^{b},
	\end{equation}
with
	\begin{equation}
		\mathfrak{h}_{ab}\left(x\right)=\eta_{MN}\frac{\partial X^{M}\left(x\right)}{\partial x^{a}}\frac{\partial X^{N}\left(x\right)}{\partial x^{b}}. \label{metric_cov}
	\end{equation}
Note we also have $\mathfrak{h}^{ab}\frac{\partial X^{M}\left(x\right)}{\partial x^{a}}\frac{\partial X^{N}\left(x\right)}{\partial x^{b}}=\eta^{MN}$.

The embedding of the 4-dimensional hypersurface into the 5-dimensional bulk can be parametrized by a constraint among coordinates $\{X^M\}$:
	\begin{equation}
		\Phi \left(X^M\right) = 0, \label{cons_X}
	\end{equation}
which is a scalar function under diffeomorphism and global Lorentz transformation of $\{X^M\}$.
Since the hypersurface is codimensional one, its normal vector (with normalization $\eta^{MN}n_M n_N =1$) is thus given by
	\begin{equation}
		n_{M} \equiv N \partial_M \Phi ,
	\end{equation}
with $\partial_M \equiv \partial \Phi/\partial X^M$ and
	\begin{equation}
		N = \frac{1}{\sqrt{\eta^{MN}\partial_{M}\Phi\partial_{N}\Phi}}.
	\end{equation}
Here the sign of $n_M$ is chosen to be compatible with the fact that $n_M$ is spacelike, such that the induced metric on the 4 dimensional hypersurface
	\begin{equation}
		h_{MN} = \eta_{MN} - n_M n_N, \label{h_MN}
	\end{equation}	
is Lorentzian.

\subsection{Unitary gauge}

Up to now, the formalism is quite general. We can now fix a gauge (unitary gauge) by choosing a specified mapping between the two set of coordinates:
	\begin{equation}
		x^{a}\mapsto X^{M} \equiv \bar{X}^{M}\left(x\right),
	\end{equation}
which corresponds to a special choice of $\{ x^{a}\}$-coordinates adapted to the embedding, i.e.,
	\begin{equation}
	\{x^a\} =\{x^{\mu},y\},\qquad \text{with}\quad \mu=0,1,2,3,
	\end{equation}
such that the metric of 5-dimensional bulk in this particular $\{ x^{a} \} $-coordinates takes the form
	\begin{eqnarray}
	\mathrm{d}s^{2} & = & \bar{\mathfrak{h}}_{ab} dx^{a}dx^{b}\nonumber \\
	& = & \bar{f}_{\mu\nu}dx^{\mu}dx^{\nu}+2N_{\mu}dx^{\mu}dy+\big(N^{2}+N_{\mu}N^{\mu}\big)dy^{2},\qquad \label{metric_ug}
	\end{eqnarray}
where $\bar{f}_{\mu\nu}$ is just the fixed induced metric on the hypersurface, which we treat as being given beforehand.
In (\ref{metric_ug}) we write $\bar{\mathfrak{h}}_{ab}$ in order to emphasize it is the expression in the unitary gauge.
It is always possible to choose $y$-coordinate to be normal to the hypersurface so that $\bar{\mathfrak{h}}_{\mu y} = N_{\mu}=0$.
Here we keep $N_{\mu}\neq 0$ for generality, while as we shall see later, all the contributions from $N_{\mu}$ drop out in the final expressions.
At this point, keep in mind that $N$ and $N_{\mu}$ must be determined by $\bar{f}_{\mu\nu}$ in order to make sure that $\bar{\mathfrak{h}}_{ab}$ is indeed describing a flat space.

In the following, we use
	\begin{equation}
		e_{a}^{M}\equiv\frac{\partial\bar{X}^{M}\left(x\right)}{\partial x^{a}},
	\end{equation}
for short.
In this unitary gauge, the normal vector to the hypersurface is given by, in $\{ X^{M}\}$-coordinate:
	\begin{equation}
		n^{M}=\frac{1}{N}\left(e_{y}^{M}-N^{M}\right), \label{nor_ug}
	\end{equation}
and in $\{x^a\}$-coordinates:
	\begin{equation}
		n_{a}\equiv\left\{ 0,\cdots,0,n_{y}\right\} =\left\{ 0,\cdots,0,N\right\} .
	\end{equation}
The induced metric on the hypersurface (\ref{h_MN}) thus becomes
	\begin{equation}
		h_{ab} = \mathfrak{h}_{ab}- n_a n_b. \label{h_ind}
	\end{equation}

\subsection{Goldstone modes on the hypersurface}

Deviation from the unitary gauge can be achieved by choosing a different mapping $x^a\mapsto X^M$, which is parametrized by
	\begin{equation}
		X^{M}\left(x\right)=\bar{X}^{M}\left(x\right)-\pi^{M}\left(x\right), \label{goldstone}
	\end{equation}
where $\pi^{M}(x)$ are five Goldstone modes in the 5-dimensional bulk.
Note (\ref{goldstone}) is simply a copy of (\ref{naive}).
Now comes the crucial point. Since the 4-dimensional hypersurface is treated as being fixed and thus nondynamical, all gauge choices should satisfy the constraint (\ref{cons_X}). That is, we must have
	\begin{equation}
		\Phi\left(\bar{X}^{M}\left(x\right)-\pi^{M}\left(x\right)\right)=0, \label{cons_pi}
	\end{equation}
which eleminates one degree of freedom among the five Goldstone modes $\pi^{M}$. Thus we are left with only four gauge modes, as is expected. This is different from the case where the hypersurface is dynamical (e.g. galileons from induced metric \cite{deRham:2010eu,Hinterbichler:2010xn,Burrage:2011bt,Goon:2011qf,Goon:2011uw}), where the position of the hypersurface becomes a dynamical variable, which cannot be gauged away.

\begin{figure}[tbp]
	\centering
	\includegraphics[width=7cm]{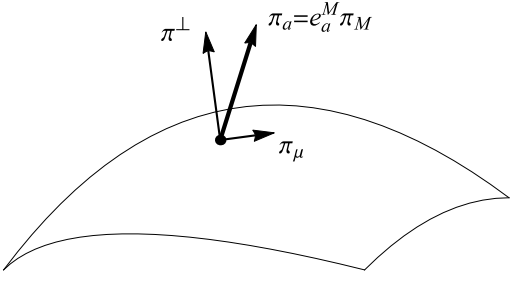}
	\caption{Illustration of the embedding of the 4-dimensional hypersurface into the 5-dimensional Minkowski background and the definition of Goldstone modes.}
	\label{fig:ebd}
\end{figure}

The components of $\pi_{M}$ in $\{x^a\}$-coordinate read
	\begin{equation}
		\pi_a \equiv \eta_{MN} e^M_a \pi^N = \{\pi_{\mu}, \pi_y\}.
	\end{equation}
Equivalently, we may write
	\begin{equation}
		\pi^M \equiv e^M_a \pi^a = e^M_{\mu} \pi^{\mu} + e^M_y \pi^{y}.
	\end{equation}	
For later convenience, we also define
	\begin{equation}
		\pi^{\perp}\equiv n_{M}\pi^{M}=n_{a}\pi^{a}=n_{y}\pi^{y}\equiv N\pi^{y}. \label{piperp}
	\end{equation}
From the 4-dimensional point of view, it may be convenient to choose the four Goldstone modes to be $\pi_{\mu}$, i.e., the direct projection of $\pi^{M}$ on the hypersurface. 	
The crucial point is, $\pi^{\perp}$ must be determined by $\pi^{\mu}$ through the constraint (\ref{cons_pi}) as 
	\begin{equation}
		\pi^{\perp}=\pi^{\perp}\left(\pi^{\mu}\right),
	\end{equation}
which is a scalar under $\{x^{\mu}\}$ reparametrization.
The embedding of the 4-dimensional hypersurface as well as the definition of 5-dimensional and 4-dimensional Goldstone modes are illustrated in Fig.\ref{fig:ebd}.

In general, we may solve $\pi^{\perp}$ in terms of $\pi^{\mu}$ perturbatively.
To this end, we expand (\ref{cons_pi}) around $\Phi\left(\bar{X}\right)=0$, which yields
	\begin{equation}
		\sum_{n=1}\left(-1\right)^{n}\frac{\partial^{n}\Phi}{\partial\bar{X}^{M_{1}}\cdots\partial\bar{X}^{M_{1}}}\pi^{M_{1}}\cdots\pi^{M_{n}}=0. \label{cons_exp}
	\end{equation}
In $\{x^a\}$-coordinate, the constraint (\ref{cons_exp}) becomes
	\begin{equation}
		\sum_{n=1}\left(-1\right)^{n}\Phi_{;a_{1}\cdots a_{n}}\pi^{a_{1}}\cdots\pi^{a_{n}}=0, \label{cons_xa}
	\end{equation}
where
	\begin{equation}
		\Phi_{;a_{1}\cdots a_{n}}\equiv e_{a_{1}}^{M_{1}}\cdots e_{a_{n}}^{M_{n}}\frac{\partial^{n}\Phi}{\partial\bar{X}^{M_{1}}\cdots\partial\bar{X}^{M_{1}}}\equiv\nabla_{a_{1}}\cdots\nabla_{a_{n}}\Phi.
	\end{equation}
Note $\Phi_{;a_1\cdots a_n}$ is symmetric since the 5-dimensional bulk is essentially flat.

By definition,
	\begin{equation}
		\nabla_{a}\Phi=\frac{1}{N}n_{a}, \label{d1_dec}
	\end{equation}
and a further derivative yields
	\begin{equation}
		\nabla_{a}\nabla_{b}\Phi=-\frac{1}{N}\left(n_{a}n_{b}\rho-2n_{(a}a_{b)}-K_{ab}\right), \label{d2_dec}
	\end{equation}
with shorthands
	\begin{equation}
		\rho\equiv \pounds_{\bm{n}}\ln N,\qquad a_{a}=-D_{a}\ln N. \label{rho_a}
	\end{equation}
In the above, $D_{a}$ is the tangent covariant derivative (compatible with $h_{ab}$ in (\ref{h_ind})), $K_{ab}\equiv D_{b}n_{a}$ is the extrinsic curvature.
The third derivative is decomposed to be\footnote{Here (\ref{d3_dec}) and (\ref{d3_s1})-(\ref{d3_s5}) are slightly different from those in \cite{Gao:2014soa,Gao:2014fra} since now the normal to the hypersurface is spacelike.} \cite{Gao:2014soa,Gao:2014fra}
	\begin{eqnarray}
	\nabla_{a}\nabla_{b}\nabla_{c}\Phi & = & \frac{1}{N}\Big(n_{a}n_{b}n_{c}\omega+2n_{a}n_{(b}\lambda_{c)}\nonumber \\
	&  & +n_{a}\zeta_{bc}+\lambda_{a}n_{b}n_{c}+2\xi_{a(b}n_{c)}+\chi_{abc}\Big),\qquad \label{d3_dec}
	\end{eqnarray}
with
	\begin{eqnarray}
	\omega & \equiv & \rho^{2}-\pounds_{\bm{n}}\rho-2a_{a}a^{a},\label{d3_s1}\\
	\lambda_{a} & \equiv & \pounds_{\bm{n}}a_{a}-2\rho a_{a}-2K_{a}^{c'}a_{c'},\label{d3_s2}\\
	\zeta_{ab} & \equiv & 2a_{a}a_{b}-\rho K_{ab}+\pounds_{\bm{n}}K_{ab}-2K_{ac'}K_{b}^{c'},\label{d3_s3}\\
	\xi_{ab} & \equiv & a_{a}a_{b}+D_{a}a_{b}-\rho K_{ab}-K_{a}^{b'}K_{b'b},\label{d3_s4}\\
	\chi_{abc} & \equiv & a_{a}K_{bc}+2K_{a(b}a_{c)}+D_{a}K_{bc},\label{d3_s5}
	\end{eqnarray}
where $\rho$ and $a_a$ are defined in (\ref{rho_a}).
The full decomposition of $\nabla_{a}\nabla_{b}\nabla_{c}\nabla_{d}\Phi$ is rather involved and we prefer not to present it in this note.
For our purpose to solve $\pi^{\perp}$ up to the fourth order in $\pi_{\mu}$, only the purely tangent part of $\nabla_{a}\nabla_{b}\nabla_{c}\nabla_{d}\Phi$ is needed, which reads
	\begin{equation}
	\nabla_{a}\nabla_{b}\nabla_{c}\nabla_{d}\Phi \supset \frac{1}{N} \chi_{abcd},
	\end{equation}
with \cite{Gao:unpublished}
	\begin{equation}
	\chi_{abcd}=K_{ab}\zeta_{cd}+K_{ac}\xi_{bd}+K_{ad}\xi_{bc}+a_{a}\chi_{bcd}+D_{a}\chi_{bcd},
	\end{equation}
where $\zeta_{ab}$ etc are defined in (\ref{d3_s3})-(\ref{d3_s5}).

Supposing that $\pi^{\perp}$ can be solved in terms of $\pi_{\mu}$ perturbatively as
	\begin{equation}
		\pi^{\perp}=\sum_{n=1} \pi_{(n)}^{\perp},
	\end{equation}
where $\pi_{(n)}^{\perp} \sim \mathcal{O}((\pi_{\mu})^n)$.
Plugging (\ref{d1_dec}), (\ref{d2_dec}) and (\ref{d3_dec}) into (\ref{cons_xa}) and using the definition for $\pi^{\perp}$ and $\pi_{\mu}$, after some manipulations we have
	\begin{eqnarray}
	\pi^{\perp} & = & \frac{1}{2}\pi^{\mu}\pi^{\nu}\bar{K}_{\mu\nu}-\frac{1}{6}\pi^{\mu}\pi^{\nu}\pi^{\rho}\bar{\nabla}_{\mu}\bar{K}_{\nu\rho}\nonumber \\
	&  & +\frac{1}{24}\pi^{\mu}\pi^{\nu}\pi^{\rho}\pi^{\sigma}\left(3\bar{K}_{\mu\nu}\bar{K}_{\rho\lambda}\bar{K}_{\sigma}^{\lambda}+\bar{\nabla}_{\mu}\bar{\nabla}_{\nu}\bar{K}_{\rho\sigma}\right)\nonumber \\
	&  & +\mathcal{O}\left((\pi_{\mu})^5\right). \label{piperp_pimu}
	\end{eqnarray}
(\ref{piperp_pimu}) explicitly depends on the extrinsic curvature $\bar{K}_{\mu\nu}$ and its derivatives.
While from the 4-dimensional point of view, the fiducial metric should not ``know'' anything about the embedding.
As we shall see in the next section, in the final expressions for the St\"{u}ckelbergized fiducial metric $f_{\mu\nu}$ (\ref{f_ind}), all the dependence on the extrinsic curvature gets suppressed after using the Gauss relation and a simple field redefinition.

\section{Covariant expansion of the fiducial metric} \label{sec:expan}

We are now ready to expand the ``St\"{u}ckelbergized'' fiducial metric given by
	\begin{equation}
		f_{\mu\nu} = \mathfrak{h}_{\mu\nu}
	\end{equation}
with $\mathfrak{h}_{ab}$ given in (\ref{metric_cov}), in terms of $\pi_{\mu}$ as well as 4-dimensional curvature quantities.
Expanding around the unitary gauge by plugging (\ref{goldstone}) into (\ref{metric_cov}), we get
	\begin{equation}
		f_{\mu\nu}=\bar{f}_{\mu\nu}-2\eta_{MN}\partial_{(\mu}\bar{X}^{M}\partial_{\nu)}\pi^{N}+\eta_{MN}\partial_{\mu}\pi^{M}\partial_{\nu}\pi^{N}, \label{cfm_ori}
	\end{equation}
which is exact. 
Note in deriving (\ref{cfm_ori}), we used $\eta_{MN}\partial_{\mu}\bar{X}^{M}\partial_{\nu}\bar{X}^{N}\equiv\bar{f}_{\mu\nu}$.
Our purpose is to rewrite (\ref{cfm_ori}) in a covariant manner in terms of 4 dimensional quantities.

For the second term in (\ref{cfm_ori}), it is easy to show that (see (\ref{dA_hh}))
	\begin{eqnarray}
	\eta_{MN}\partial_{\mu}\bar{X}^{M}\partial_{\nu}\pi^{N} & = & e_{\mu}^{M}e_{\nu}^{N}\partial_{N}\pi_{M}\nonumber \\
	& \equiv & \bar{\nabla}_{\nu}\pi_{\mu}+\bar{K}_{\mu\nu}\pi^{\perp},\label{cov_pi}
	\end{eqnarray}
where $\pi^{\perp}$ is defined in (\ref{piperp}).
For the last term in (\ref{cfm_ori}), first we have
	\begin{eqnarray}
	&  & \eta_{MN}\partial_{\mu}\pi^{M}\partial_{\nu}\pi^{N}\nonumber \\
	& = & \big(\bar{\mathfrak{h}}^{ab}e_{a}^{M}e_{b}^{N}\big)e_{\mu}^{M'}e_{\nu}^{N'}\partial_{M'}\pi_{M}\partial_{N'}\pi_{N}\nonumber \\
	& = & \bar{\mathfrak{h}}^{\rho\sigma}\big(e_{\rho}^{M}e_{\mu}^{M'}\partial_{M'}\pi_{M}\big)\big(e_{\sigma}^{N}e_{\nu}^{N'}\partial_{N'}\pi_{N}\big)\nonumber \\
	&  & +\bar{\mathfrak{h}}^{\rho y}\big(e_{\rho}^{M}e_{\mu}^{M'}\partial_{M'}\pi_{M}\big)\big(e_{y}^{N}e_{\nu}^{N'}\partial_{N'}\pi_{N}\big)\nonumber \\
	&  & +\bar{\mathfrak{h}}^{y\sigma}\big(e_{y}^{M}e_{\mu}^{M'}\partial_{M'}\pi_{M}\big)\big(e_{\sigma}^{N}e_{\nu}^{N'}\partial_{N'}\pi_{N}\big)\nonumber \\
	&  & +\bar{\mathfrak{h}}^{yy}\big(e_{y}^{M}e_{\mu}^{M'}\partial_{M'}\pi_{M}\big)\big(e_{y}^{N}e_{\nu}^{N'}\partial_{N'}\pi_{N}\big),\label{cov_pi2_ori}
	\end{eqnarray}
where $\bar{\mathfrak{h}}^{\mu\nu}$ etc are the components of the matrix inverse of $\bar{\mathfrak{h}}_{ab}$ in (\ref{metric_ug}).
While using (\ref{nor_ug}), we have
	\begin{eqnarray}
	&  & e_{y}^{M}e_{\mu}^{M'}\partial_{M'}\pi_{M}\nonumber \\
	& = & Nn^{M}e_{\mu}^{M'}\partial_{M'}\pi_{M}+N^{\rho}e_{\rho}^{M}e_{\mu}^{M'}\partial_{M'}\pi_{M}\nonumber \\
	& \equiv & N\left(\bar{\nabla}_{\mu}\pi^{\perp}-\bar{K}_{\mu}^{\rho}\pi_{\rho}\right)+N^{\rho}\left(\bar{\nabla}_{\mu}\pi_{\rho}+\bar{K}_{\mu\rho}\pi^{\perp}\right), \label{dpi_ymu}
	\end{eqnarray}
where in the last step we used (\ref{dA_hn}).
Plugging (\ref{cov_pi}) and (\ref{dpi_ymu}) into (\ref{cov_pi2_ori}), simple manipulation yields
	\begin{eqnarray}
	&  & \eta_{MN}\partial_{\mu}\pi^{M}\partial_{\nu}\pi^{N}\nonumber \\
	& = & \bar{g}^{\rho\sigma}\left(\bar{\nabla}_{\mu}\pi_{\rho}+\bar{K}_{\mu\rho}\pi^{\perp}\right)\left(\bar{\nabla}_{\nu}\pi_{\sigma}+\bar{K}_{\nu\sigma}\pi^{\perp}\right)\nonumber \\
	&  & +\left(\bar{\nabla}_{\mu}\pi^{\perp}-\bar{K}_{\mu}^{\rho}\pi_{\rho}\right)\left(\bar{\nabla}_{\nu}\pi^{\perp}-\bar{K}_{\nu}^{\sigma}\pi_{\sigma}\right),\label{cov_pi2}
	\end{eqnarray}
where the dependence on $N$ and $N_{\mu}$ exactly gets cancelled.
Putting all the above together, finally we write
	\begin{eqnarray}
	f_{\mu\nu} & = & \bar{f}_{\mu\nu}-2\bar{\nabla}_{(\mu}\pi_{\nu)}-2\bar{K}_{\mu\nu}\pi^{\perp}\nonumber \\
	&  & +\left(\bar{\nabla}_{\mu}\pi_{\rho}+\bar{K}_{\mu\rho}\pi^{\perp}\right)\left(\bar{\nabla}_{\nu}\pi^{\rho}+\bar{K}_{\nu}^{\rho}\pi^{\perp}\right)\nonumber \\
	&  & +\left(\bar{\nabla}_{\mu}\pi^{\perp}-\bar{K}_{\mu}^{\rho}\pi_{\rho}\right)\left(\bar{\nabla}_{\nu}\pi^{\perp}-\bar{K}_{\nu}^{\sigma}\pi_{\sigma}\right).\label{cfm}
	\end{eqnarray}
(\ref{cfm}) is the one of the main results in this work.
Keep in mind that $\pi^{\perp}$ is a function of $\pi_{\mu}$ given in (\ref{piperp_pimu}).

Now our task is to plug the expression for $\pi^{\perp}$ in terms of $\pi^{\mu}$ (\ref{piperp_pimu}) into (\ref{cfm}) in order to derive ``covariant'' expansion for $f_{\mu\nu}$ in terms of $\pi^{\mu}$ and 4-dimensional geometric quantities.
We may write
	\begin{equation}
		f_{\mu\nu}-\bar{f}_{\mu\nu}=\sum_{n=1}f_{\mu\nu}^{(n)\mathrm{Ebd}},
	\end{equation}
where $f_{\mu\nu}^{(n)\mathrm{Ebd}}\sim\mathcal{O}\left(\left(\pi_{\mu}\right)^{n}\right)$ and superscript ``Ebd'' stands for ``embedding''.
After some manipulations, we have, at the linear order,
	\begin{equation}
		f_{\mu\nu}^{(1)\mathrm{Ebd}}=-2\bar{\nabla}_{(\mu}\pi_{\nu)}, \label{f1_fin}
	\end{equation}
and at the quadratic order,
	\begin{eqnarray}
	f_{\mu\nu}^{(2)\mathrm{Ebd}} & = & \bar{\nabla}_{\mu}\pi_{\rho}\bar{\nabla}_{\nu}\pi^{\rho}-\left(\bar{K}_{\mu\nu}\bar{K}_{\rho\sigma}-\bar{K}_{\mu\rho}\bar{K}_{\nu\sigma}\right)\pi^{\rho}\pi^{\sigma}\nonumber \\
	& = & \bar{\nabla}_{\mu}\pi_{\rho}\bar{\nabla}_{\nu}\pi^{\rho}-\bar{R}_{\mu\rho\nu\sigma}\pi^{\rho}\pi^{\sigma},\label{f2_fin}
	\end{eqnarray}
where in the last step we used the Gauss relation
	\begin{equation}
		\bar{K}_{\mu\nu}\bar{K}_{\rho\sigma}-\bar{K}_{\mu\rho}\bar{K}_{\nu\sigma}=\bar{R}_{\mu\sigma\nu\rho}, \label{gauss}
	\end{equation}
since the 5 dimensional bulk is flat.
Note $f^{(1)\mathrm{Ebd}}_{\mu\nu}$ in (\ref{f1_fin}) and $f^{(2)\mathrm{Ebd}}_{\mu\nu}$ in (\ref{f2_fin}) exactly coincides with the results got in the RNC approach $f^{(1)\mathrm{RNC}}_{\mu\nu}$ in (\ref{f1_RNC}) and $f^{(2)\mathrm{RNC}}_{\mu\nu}$ in (\ref{f2_RNC}) respectively.

At the cubic order,
	\begin{eqnarray}
	f_{\mu\nu}^{(3)\mathrm{Ebd}} & = & \frac{1}{3}\pi^{\alpha}\pi^{\beta}\pi^{\rho}\left(\bar{K}_{\mu\nu}\bar{\nabla}_{\alpha}\bar{K}_{\beta\rho}-3\bar{K}_{\alpha(\mu}\bar{\nabla}_{\nu)}\bar{K}_{\beta\rho}\right)\nonumber \\
	&  & -\pi^{\alpha}\pi^{\beta}\bar{\nabla}_{(\mu}\pi^{\rho}\left(2\bar{K}_{\nu)\alpha}\bar{K}_{\beta\rho}-\bar{K}_{\nu)\rho}\bar{K}_{\alpha\beta}\right).\label{f3_ori}
	\end{eqnarray}
At this point, one may be concerned about the presence of extrinsic curvature, since the final expressions should depend only on intrinsic 4 dimensional quantities.
Fortunately, by employing the Gauss relation (\ref{gauss}) many times, it is possible to recast (\ref{f3_ori}) to be
	\begin{equation}
		f_{\mu\nu}^{(3)\mathrm{Ebd}}=f_{\mu\nu}^{(3)\mathrm{RNC}}-\frac{1}{3}\bar{\nabla}_{(\mu}\left(\pi^{\alpha}\pi^{\beta}\pi^{\rho}\bar{K}_{\nu)\rho}\bar{K}_{\alpha\beta}\right), \label{f3_fin}
	\end{equation}
where $f^{(3)\mathrm{RNC}}_{\mu\nu}$ is given in (\ref{f3_RNC}).
Please note in deriving (\ref{f3_fin}), we never use $\pi^{\nu}\bar{\nabla}_{\mu}\pi^{\mu}=0$, which is the crucial assumption in the RNC approach (see Sec.\ref{sec:RNC} and \cite{Gao:2014ula} for details).
The second term on the right-hand-side of (\ref{f3_fin}), may be absorbed by a field redefinition
	\begin{equation}
		\pi_{\mu}\rightarrow \tilde{\pi}_{\mu} \equiv \pi_{\mu}+\Delta^{(3)}_{\mu}+\mathcal{O}\left((\pi_{\mu})^{4}\right), \label{redef_3}
	\end{equation}
with
	\begin{equation}
		\Delta_{\mu}^{(3)}=\frac{1}{6}\pi^{\alpha}\pi^{\beta}\pi^{\rho}\bar{K}_{\mu\rho}\bar{K}_{\alpha\beta}. \label{Delta3}
	\end{equation}
This can be seen easily since $f^{(1)\mathrm{RNC}}_{\mu\nu}(\Delta_{\rho}^{(3)})$ will exactly reproduce the second term on the right-hand-side of (\ref{f3_fin}).

Similarly, at the quartic order, first we got the expression for $f^{(4)\mathrm{Ebd}}_{\mu\nu}$ depending on the extrinsic curvature
\begin{widetext}
	\begin{eqnarray}
	f_{\mu\nu}^{(4)\mathrm{Ebd}} & = & -\frac{1}{12}\pi^{\alpha}\pi^{\beta}\pi^{\rho}\pi^{\sigma}\Big[3\bar{K}_{\alpha\beta}\left(\bar{K}_{\mu\nu}\bar{K}_{\rho\lambda}\bar{K}_{\sigma}^{\lambda}-\bar{K}_{\mu\lambda}\bar{K}_{\nu}^{\lambda}\bar{K}_{\rho\sigma}\right)\nonumber \\
	&  & \qquad\qquad\qquad\quad+\bar{K}_{\mu\nu}\bar{\nabla}_{\alpha}\bar{\nabla}_{\beta}\bar{K}_{\rho\sigma}-3\bar{\nabla}_{\mu}\bar{K}_{\alpha\beta}\bar{\nabla}_{\nu}\bar{K}_{\rho\sigma}-4\bar{K}_{\sigma(\mu}\bar{\nabla}_{\nu)}\bar{\nabla}_{\alpha}\bar{K}_{\beta\rho}\Big]\nonumber \\
	&  & -\frac{1}{3}\pi^{\alpha}\pi^{\beta}\pi^{\rho}\bar{\nabla}_{(\mu}\pi^{\sigma}\left(\bar{K}_{\nu)\sigma}\bar{\nabla}_{\alpha}\bar{K}_{\beta\rho}-3\bar{\nabla}_{\nu)}\bar{K}_{\alpha\beta}\bar{K}_{\rho\sigma}-3\bar{K}_{\nu)\alpha}\bar{\nabla}_{\sigma}\bar{K}_{\beta\rho}\right)\nonumber \\
	&  & +\pi^{\alpha}\pi^{\beta}\bar{\nabla}_{\mu}\pi^{\rho}\bar{\nabla}_{\nu}\pi^{\sigma}\bar{K}_{\alpha\rho}\bar{K}_{\beta\sigma}.\label{f4_ori}
	\end{eqnarray}
\end{widetext}
Using the Gauss relation (\ref{gauss}) again, after some manipulations, (\ref{f4_ori}) can be recast as
	\begin{eqnarray}
	f_{\mu\nu}^{(4)\mathrm{Ebd}} & = & f_{\mu\nu}^{(4)\mathrm{RNC}}+2\bar{\nabla}_{(\mu}\pi^{\rho}\bar{\nabla}_{\nu)}\Delta_{\rho}^{(3)}-2\bar{R}_{(\mu\phantom{\rho}\nu)}^{\phantom{(\mu}\rho\phantom{\nu)}\sigma}\Delta_{\rho}^{(3)}\pi_{\sigma}\nonumber \\
	&  & -2\bar{\nabla}_{(\mu}\Delta_{\nu)}^{(4)},\label{f4_fin}
	\end{eqnarray}
where $f^{(4)\mathrm{RNC}}_{\mu\nu}$ is given  in (\ref{f4_RNC}), $\Delta^{(3)}_{\mu}$ is defined in (\ref{Delta3}), and
	\begin{equation}
		\Delta_{\mu}^{(4)}=-\frac{1}{24}\pi^{\alpha}\pi^{\beta}\pi^{\rho}\pi^{\sigma}\left(\bar{K}_{\rho\alpha}\bar{\nabla}_{\mu}\bar{K}_{\sigma\beta}+2\bar{K}_{\alpha\mu}\bar{\nabla}_{\sigma}\bar{K}_{\rho\beta}\right). \label{Delta4}
	\end{equation}
It is interesting to note that although $f^{(4)\mathrm{Ebd}}_{\mu\nu}$ itself does not coincide with $f^{(4)\mathrm{RNC}}_{\mu\nu}$, their difference can be absorbed by the following field redefinition
	\begin{equation}
		\pi_{\mu}\rightarrow\tilde{\pi}_{\mu}=\pi_{\mu}+\Delta_{\mu}^{(3)}+\Delta_{\mu}^{(4)}+\mathcal{O}\left((\pi_{\mu})^{5}\right), \label{redef_4}
	\end{equation}
which is also consisitent with (\ref{redef_3}).
That is, we have 
	\begin{equation}
		f_{\mu\nu}^{\mathrm{Ebd}}\left(\pi_{\rho}\right)=f_{\mu\nu}^{\mathrm{RNC}}\left(\tilde{\pi}_{\rho}\right),
	\end{equation}
where $\tilde{\pi}_{\mu}$ is given in (\ref{redef_4}) up to the fourth order.

\section{Conclusion} \label{sec:con}

The problem of covariant formulation for the St\"{u}ckelberg analysis with a non-flat fiducial metric (or around a general background) has been known for some time.
In \cite{Gao:2014ula} a covariant St\"{u}ckelberg expansion was developed based on the Riemann normal coordinates.
In this note we explore an alternative approach by considering the $4$-dimensional curved space being a hypersurface embedded in a $5$-dimensional Minkowski background, in which the Goldstone modes and the St\"{u}ckelberg expansion can be performed in the standard manner.
After eliminating one Goldstone modes through the constraint (\ref{cons_pi}) and then projecting onto the 4-dimensional hypersurface, we are able to expand the St\"{u}ckelbergized fiducial metric (\ref{f_ind}) in terms of 4-dimensional Goldstone modes $\pi_{\mu}$ as well as 4-dimensional geometric quantities, which are given in (\ref{f1_fin}), (\ref{f2_fin}), (\ref{f3_fin}) and (\ref{f4_fin}) respectively, up to the fourth order in $\pi_{\mu}$.
Strikingly, the two approaches (RNC and embedding), although quite different from each other, give exactly coincide results after a simple field redefinition (\ref{redef_4}).

We expect the formalism developed in this note, may shed some light on the St\"{u}ckelberg analysis as well as the decoupling limit of the massive gravity on a general background.
There are some questions left to be answered. First it is important to find a geometric meaning for the field redefinition (\ref{redef_4}), which implies that the correct ``covariant'' Goldstone modes from the 4-dimensional point of view are actually nonlinear functions $\pi_{\mu}$ (instead of $\pi_{\mu}$ themselves). 
Moreover, as being emphasized before, higher codimensions may be needed in order to embed a general curved space into a flat background. It is thus interesting to generalize the formalism in this note to such a case.


\acknowledgments

I would like to thank
Masahide Yamaguchi
for discussions and comments.
I was supported by JSPS Grant-in-Aid for Scientific Research No. 25287054.

\appendix

\section{Projection of $\nabla_a A_b$}

Let us consider a codimension-one hypersurface in an arbitrary background with metric $g_{ab}$.
The normal vector $n_a$ is normalized as $n_a n^a = +1$, which is spacelike.
The induced metric on the hypersurface is $h_{ab} = g_{ab} - n_a n_b$. 
For an arbitrary vector field $A_{a}$, we may write
	\begin{equation}
		A_{a}\equiv n_{a}A^{\perp}+A_{a}^{\parallel},
	\end{equation}
with
	\begin{equation}
		A^{\perp}=n^{a}A_{a},\qquad A_{a}^{\parallel}=h_{a}^{a'}A_{a'}.
	\end{equation}
It is easy to show that
	\begin{eqnarray}
	h_{a}^{a'}h_{b}^{b'}\nabla_{a'}A_{b'} & = & D_{a}A_{b}^{\parallel}+K_{ab}A^{\perp},\label{dA_hh}\\
	h_{a}^{a'}n^{b'}\nabla_{a'}A_{b'} & = & D_{a}A^{\perp}-K_{a}^{b}A_{b}^{\parallel},\label{dA_hn}
	\end{eqnarray}
where $D_a$ is the covariant derivative compatible with $h_{ab}$, $K_{ab}\equiv D_a n_b$ is the extrinsic curvature.


\bibliography{Gao}

\end{document}